# New Kids on the Block: On the impact of information retrieval on contextual resource integration patterns

*Extended Abstract*

**Martin Semmann**
Universität Hamburg, Germany
martin.semmann@uni-hamburg.de

**Mahei Manhai Li**
University of Kassel, Germany
University of St.Gallen, Switzerland
mahei.li@uni-kassel.de

## Introduction

The rise of new modes of interaction with AI skyrocketed the popularity, applicability, and amount of use cases. Despite this evolution, conceptual integration is falling behind. Studies suggest that there is hardly a systematization in using AI in organizations (Uba et al., 2023). Thus, by taking a service-dominant logic perspective, specifically, the concept of resource integration patterns, the most potent application of AI for organizational use - namely information retrieval - is analyzed. In doing so, the paper proposes a systematization that can be applied to deepen understanding of core technical concepts, further investigate AI in contexts, and help explore research directions guided by SDL.

With the public unveiling of openAI in November 2022, generative AI (genAI) has garnered much attention in both academic (Böhmann et al., 2023; Wessel et al., 2023) and industry circles (McKinsey, 2023). Recent studies on the potential of genAI, which largely rely on large language model (LLM) systems, ranging from automatization (Engel et al., 2023) to improving knowledge work (Anthony et al., 2023; Dell'Acqua et al., 2023) to creating novel business models (Kanbach et al., 2023). However, with the generation of novel content, LLM users are at risk of being presented with factually incorrect data – so-called hallucinations (Blom, 2010), which are hard to discern from factually correct ones (Jia et al 2023). Researchers have turned toward two different research approaches:

The first approach to address the hallucination problem is to increase output by integrating the human actor into the generative AI system. For example, the hallucination detection task can be directly sourced to the human end-users. This means that explainable AI service elements are presented in tandem with the generative AI system output to help users discern its factual validity (e.g. (Kadavath et al., 2022; Manakul et al., 2023). Another example approach focuses on the alterations of prompts via different instruction methods, such as zero-shot or few-shot learning, or by applying different contexts and leading the generative AI system toward a desirable output through chaining (Chase, 2022/2022).

The second approach focuses on different retrieval-augmented generation types to improve the factuality of gen-AI outputs (e.g.: (Asai et al., 2023; Jiang et al., 2023)). One viable approach relies on integrating other subsystems that are not inherently part of the LLM, such as by deploying knowledge graphs (Martino et al., 2023). Knowledge graphs are factual knowledge representations and have long been used to assure the content quality of Q&A chatbots (Ait-Mlouk & Jiang, 2020; P. Sukhwal et al., 2022; P. C. Sukhwal et al., 2023) or to discover and retrieve information (Banerjee, Yimam, et al., 2023). Recently, KGs have been deployed to influence generative AI systems toward providing factual assurance (Ji et al., 2023). Similarly, the integration of traditional databases has been a reliable technology over the last decades as long the database content has been thoroughly vetted for data quality measures (Imielinski & Mannila, 1996; Sarawagi, 2007).

While both approaches are forms of hybrid intelligence, in which humans and AI systems interact to create superior results (Dellermann et al., 2019), we call our attention to different retrieval-augmented generation structures. Service dominant logic provides us with a theoretical framework to study these structures as novel resource integration patterns that leverage existing externalized knowledge and its underlying knowledge structure (Lusch & Vargo, 2006; Vargo, 2008; Vargo et al., 2017). Our paper will outline several such resource integration patterns.

From an information retrieval perspective, our proposed resource integration patterns stress the importance of providing additional context information to improve the value of each retrieved or (partially) generated information. It highlights the synergetic effects of generative AI (LLMs) and information retrieval models (Jiang et al., 2023). Based on a service perspective, our presented resource integration patterns are a novel instantiation of the service concept *value-in-context* (Chandler & Vargo, 2011). Hence, we argue that a GenAI system's value realization is strongly determined by its context information. As such, we argue that value-in-context should be further classified to explicitly include information context that augments information output (e.g., factually known data to enhance LLM-generated information). Further investigations into the concept of value-in-context could reveal the nuances of contextual resource integration patterns.

From a service perspective, we observe the phenomena of novel resource integration patterns for LLM-based systems. Novel solutions like ChatGPT that proposes an interface that is easily approachable by internal as well as external users consequently lead to new opportunities to design services, to transform value creation, and especially, to reduce efforts needed in knowledge-intense, person-oriented services (Menschner et al., 2011). The potential lever on information retrieval is accordingly potent, as complex informational needs can – with several limitations – be fulfilled while further experience and or expert knowledge is not necessarily needed.

## Resource Integration Patterns of Information Retrieval

Within service systems, a major conceptual foundation is the integration of resources as a matter of realizing value. Such mechanisms of integration are subsumed as resource integration patterns that enable a systematized (re-)usability within different instantiations of value-in-context (Chandler & Vargo, 2011). Following this conceptual foundation, the application of artificial intelligence, especially generative AI, can be understood as a novel approach to integrating resources. Information retrieval is a basic and widespread application of RIPs in service science and a technological use case for generative artificial intelligence. Thus, we combine these research directions and propose archetypes of information retrieval as symbolic representations of RIPs.

First, basic information retrieval with the utilization of information systems can be seen as traditional storage of data in databases accessible via specific languages, i.e., SQL, to enable a systematic approach to retrieve data and allow basic operations based on the dataset (Figure 1).

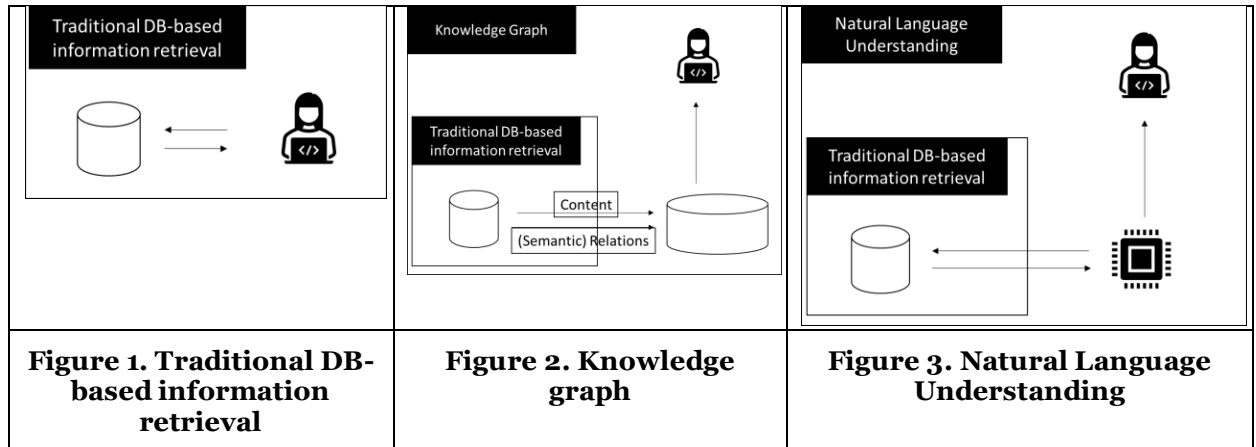

| **Figure 1. Traditional DB-based information retrieval** | **Figure 2. Knowledge graph** | **Figure 3. Natural Language Understanding** |

Second, knowledge graphs are a more recent approach to represent data and store relational information of the data. Thus, enabling a broader perspective on data that provides a better understanding of the domain and perspective on the datasets. The information retrieval can leverage the semantic relationship of pre-defined data models to increase the information value for the user. The semantic relationships are extracted and saved within a knowledge graph. The knowledge graph can be instantiated as relational databases, while some technology providers have started to provide graph-based databases better to accommodate the technological requirements of knowledge-graph information retrieval activities. Consequently, a more encompassing perspective on informational needs can be retrieved.

The third RIP can be represented as natural language understanding (Ait-Mlouk & Jiang, 2020). Natural language understanding services typically encompass conversational agents' core text mining technologies. It is a collection of services, often accessible by APIs, which allows other systems to analyze qualitative texts and recognize intents. Information retrieval systems handle either structuring intents based on its semantic properties or providing the necessary information to the governing system on post-intent activities, e.g., by providing requested information in Q&A systems. For example, a conversational Q&A system can interpret the user request by identifying the related intent, which would trigger the retrieval of pre-structured data.

Fourth, LLMs represent a novel approach to information retrieval. In contrast to traditional extractive tasks, LLMs are often employed for generative tasks. These language models are pre-trained systems, often on large datasets, and fine-tuned via reinforcement learning and human-in-the-loop to provide a human-like conversational interface (e.g. (Ouyang et al., 2022). For users, value is realized by accessing an interactive system that provides necessary information through extractive or generative tasks. Fine-tuning a language model to the intended purpose by providing the necessary contextual information is often done via supervised instructions and chats to leverage high-quality, low-quantity data to improve the model. The contextual information leads to improved models and more useful system output.

Lastly, Retrieval Augmented Generation is a combination of NLU and LLM that interact to enable information retrieval based on factual data – NLU –used to generate responses with the LLM (Fatemi et al., 2023). Thus, chains of recursive prompts interacting between both systems can be achieved to reduce hallucinations while acting upon contextual data. They typically work following a retrieve-then-work paradigm, where relevant contextual information is found and retrieved from external sources and is followed by another generation system, which is conditioned on both the retrieved contextual information and the user input to provide the augmented information to the end-user (Karpukhin et al., 2020). The knowledge graphs are often used to prompt the LLMs to improve output (Qi et al., 2023) and multiple iterative interactions between RAGs, knowledge graphs, and further fine-tuning systems to accommodate iterative improvement during output generation (e.g. (Trivedi et al., 2022; Yao et al., 2022)).

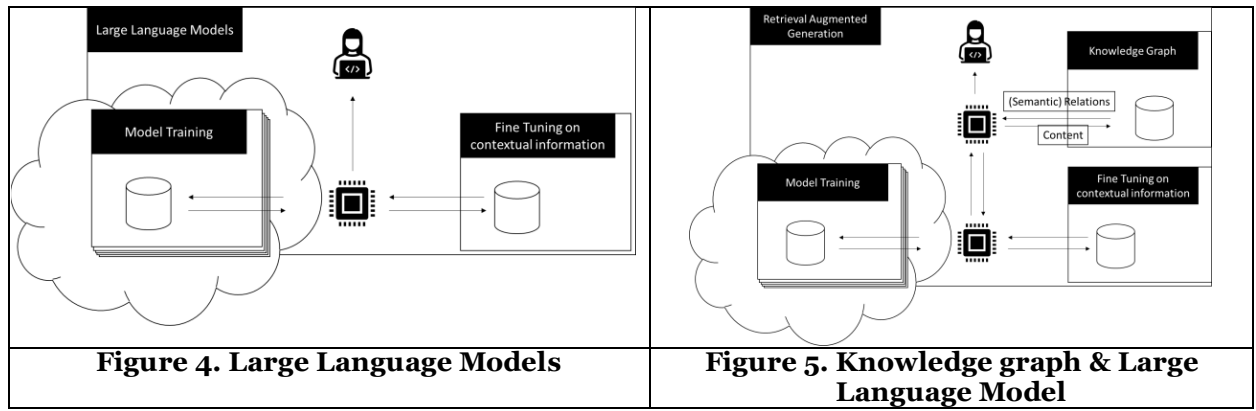

| **Figure 4. Large Language Models** | **Figure 5. Knowledge graph & Large Language Model** |

# Resulting Research Directions for Service Science

This set of conceptual resource integration patterns can and should be used as a vantage point for future research endeavors. Especially the potentials of NLU, LLM, and RAG are especially manifold, and there is a lack of research on it. From a technical perspective, the integration in service architectures need to be investigated. From a socio-technical perspective, we need to investigate further how human-computer interaction changes and what potentials occur. This includes but extends the ongoing research on conversational agents (Banerjee, Poser, et al., 2023; Peters et al., 2016). From a service perspective, to which we want to emphasize that the impact on core foundations needs to be assessed and further developed. Reflecting service dominant logic, the emergence of NLU, LLM, and RAG has a profound impact. As those RIPs are building on what is traditionally coined as operant resources, like documents or, more generally, resources upon which actions are taken, the RIPs themselves can be considered operant as well as operand resources. Basically, the RIPs are technical solutions that do not bring in experience or skills that would deem them as operand resources. On the other hand, as the RIPs autonomously act upon operand resources, namely analyze, aggregate, and summarize operant resources, the differentiation between those categories of resources needs to be further explored. Consequently, the impact on value creation needs to be systematically assessed. Even more, nested architectures of the RIPs lead to novel agents that can be dynamically generated and deployed (OpenAI, 2023).

Considering value-creation, the potential to automatically enhance contextual awareness and applicability, the discourse on the concept of realization of value and its dependence on context should be revisited (Chandler & Vargo, 2011). The dominant conceptualization of value realization considers the time as core anchor in value-in-use (Vargo et al., 2017). In contrast, the concept of value-in-context (Chandler & Vargo, 2011) is more inclusive and integrates the broader context while realizing the value explicitly. However, our resource integration patterns propose a more nuanced perspective on value-in-context by focusing on the interrelationship of training data, language models, and human input during training and system use. We focus on an ongoing dialectic between value-in-context (model training) and value-in-use (system use). Our resource integration patterns leverage context information to improve the chances of providing viable information to end-users. As such, we propose that our resource integration patterns form a class that recognizes that value-in-context can be increased before its realization (value-in-use). Furthermore, input from users during use are also leveraged to improve LLM output, either directly (by means of LLM reinforcement learning based on prompts – e.g., Zhang et al.2023e) or indirectly (e.g., by means of further system behavior analytics). This stresses the importance to distinguish between different value realization mechanisms and can guide future value co-creation applications.

As the novel RIPs bear the potential for flexibilization and to contextualize information retrieval, such a broader understanding of value realization could enable researchers to understand the underlying processes and constructions better. State-of-the-art Q&A bots apply information retrieval resource integration patterns that integrate cognitive discerning capabilities pre-generation and post-generation (e.g. (Gao et al., 2023)). The role of human actors within our resource integration patterns also calls for further research.

## Conclusion

The research presented can be understood as an initial vantage point to systematize the discourse in service science to understand the changes in value-creation imposed by artificial intelligence. The potential to significantly affect the organizational practices alone in information retrieval is manifold. Simultaneously, a conceptualization with a solid foundation in service science is missing. Thus, we propose synthesizing the technical architecture that can be (re-)combined as resource integration patterns. This abstraction supports practice and research to investigate use cases systematically and to give a guiding set of representations.